# Supercontinuum generation without residual pump peak through multiple coherent pump seeds

**DONG QIU,**[1] **WEI LIN,**[2] **HONG-JIE CHEN,**[1] **WEN-JING LIAO,**[1] **WEI-YI HONG,**[1] **HONG-ZHAN LIU,**[1] **WEI-CHENG CHEN,**[3,4] **HU CUI,**[1] **ZHI-CHAO LUO,**[1] **WEN-CHENG XU,**[1] **AND AI-PING LUO,**[1,*]

[1] *Guangdong Provincial Key Laboratory of Nanophotonic Functional Materials and Devices, South China Normal University, Guangzhou, Guangdong 510006, China*
[2] *State Key Laboratory of Luminescent Materials and Devices, South China University of Technology, Guangzhou, Guangdong 510640, China*
[3] *School of Physics and Optoelectronic Engineering, Foshan University, Foshan, Guangdong 528000, China*
[4] *chenwch@fosu.edu.cn*
*\* [luoaiping@scnu.edu.cn](luoaiping@scnu.edu.cn)*

**Abstract:** Residual pump peak in fiber-based supercontinuum, as a general phenomenon, limits its practical application. We report a novel supercontinuum generation (SCG) in a conventional highly nonlinear fiber (HNLF) through multiple coherent pump technique, which eliminates the residual pump peak existed in conventional SCG. The multiple coherent pump technique is realized by double bound-state solitons achieved from a homemade mode-locked fiber laser. We further compare the SCGs pumped by conventional bound-state soliton and single soliton. It confirms that the effective elimination of the residual pump peak in supercontinuum owes to higher transferring efficiency of the pump energy to new generated frequencies in the multiple coherent pump scheme. The use of multiple coherent pump scheme, i.e., double bound-state solitons, provides a new, simple and promising method to obtain flat supercontinuum source.

## 1. Introduction

The supercontinuum sources attract a great deal of attention because of their important applications in diverse fields such as optical communication [1], frequency-comb [2], serial time-encoded amplified microscopy (STEAM) [3] and optical coherence tomography (OCT) [4]. Supercontinuum generation (SCG), firstly reported in bulk medium by Alfano and Shapiro [5], is known as a particular spectral broadening process in a nonlinear medium pumped by high-power pulse or continuous wave (CW). SCG research further extended to optical fibers in 1976 [6]. The observed spectral broadening was attributed to cascaded stimulated Raman scattering (SRS) and self-phase modulation (SPM) effects. After that, SCGs in fibers were widely investigated [7-12]. In order to obtain broadband supercontinuum, researchers adopted various pump seeds from soliton pulse [8], Q-switched pulse [13], noise-like pulse [11,14] to dissipative soliton resonance [15], and special fibers from ZBLAN fiber [16-18] to photonic crystal fiber based on $As_2Se_3$ chalcogenide glass [19]. The supercontinuum could cover the wavelengths from ultraviolet to infrared spectra [7,12,15-19], bringing wider applications. However, it is also found that the generated supercontinuum is not flat due to the existence of a residual spectral peak at the pump wavelength [10,13,15-17], which should be taken out in practical applications. Up to date, few techniques have been put forward, such as cascading specific optical fibers [20] and designing special photonic crystal fibers [21], to remove the residual pump peak in supercontinuum. But they undoubtedly increase the cost and complexity of the supercontinuum sources. Thus, the elimination of residual pump peak in SCG is always an interesting and significant issue. It is necessary to propose an effective solution for eliminating the residual pump peak in supercontinuum without any special components and achieving a flat spectrum from a simple system.

In this work, we address this issue of how to remove the residual pump peak in SCG by adopting the multiple coherent pump mechanism in a conventional highly nonlinear fiber (HNLF). The pump seeds we employed are double bound-state solitons, achieved from a fiber laser mode-locked through nonlinear polarization rotation (NPR) technique. The obtained supercontinuum realizes the elimination of the residual pump peak due to multiple coherent pump seeds, which improve the transferring efficiency of the energy at the pump to new generated frequencies. Moreover, for comparison, we employ another two types of pulses as seed pulses, namely, conventional bound-state soliton and single soliton. It is found that in these two cases, the generated supercontinuums still keep strong residual pump peak. The double bound-state solitons served as the seed pulses of the multiple coherent pump scheme offer a simple, fresh and promising approach to obtain flat supercontinuum.

## 2. Experimental setup

Figure 1 depicts the schematic of the supercontinuum source. It consists of three parts: a mode-locked fiber ring laser acted as the seed source, a homemade erbium-doped fiber amplifier (EDFA) for amplifying the seed pulse and an 85-m long HNLF (YOFC, NL1016-B) with nonlinear coefficient of 10 $W^{-1}km^{-1}$ for broadening the pulse spectrum. The seed pulse laser is a typical fiber laser mode-locked by NPR technique, including a 5-m long erbium-doped fiber (EDF, Corning Er1550C3CT VHA) pumped by a 980 nm laser diode (LD) through a wavelength-division-multiplexer (WDM), two polarization controllers (PCs), a polarization-dependent isolator (PD-ISO) and an output coupler (OC) with 30% output. The spectral and temporal characteristics of the output pulse are monitored by an optical spectrum analyzer (Yokogawa, AQ6375B) and a real-time oscilloscope (Tektronix, DSA70804) with a photodiode detector (NewFocus, P818-BB-35F, 12.5 GHz), respectively. The fine structure of the laser pulse is further measured with an autocorrelator (Femtochrome, FR-103WS). Then the output pulse as the seed pulse is transmitted into the EDFA, comprising a 2-m long EDF (Nufern SM-ESF-7/125) which is also pumped by a 980 nm LD through a WDM with a pump protector (PP) for improving the power. Finally, the amplified pulse is injected into the HNLF to generate supercontinuum. Note that for monitoring the pulse power input in the HNLF, we add an optical coupler to extract 1% light and measure it with an optical power meter (OPM).

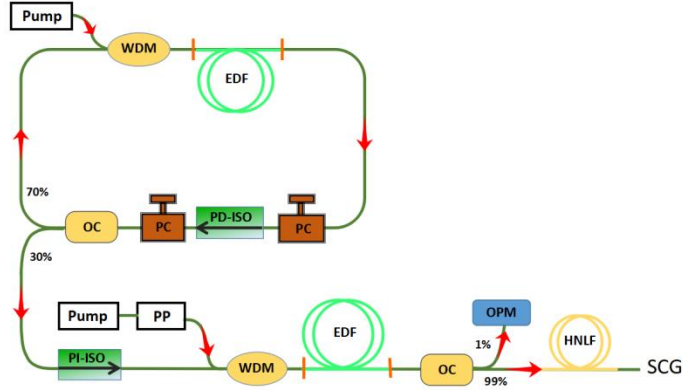

Fig. 1. Schematic of the proposed supercontinuum source.

## 3. Results and discussions

### 3.1 Supercontinuum generation with double bound-state solitons

By tuning the pump power and carefully manipulating the PCs in the seed pulse laser, we obtained the double bound-state solitons. The typical properties of the pulses are depicted in Fig. 2. From Fig. 2(a), we see a modulation with a period of 0.68 nm across the spectrum with a central wavelength of 1567 nm. For better clarity, we also show the spectrum with a small range in the inset of Fig. 2(a). The regular spectral modulation is one typical characteristic of bound-state soliton [22,23]. The corresponding pulse-train is presented in Fig. 2(b). The fundamental repetition rate is 13.2 MHz, which is determined by the cavity length. To confirm that the pulse is bound-state soliton, we measured its autocorrelation trace, as demonstrated in Fig. 2(c). Note that there are five pairs of solitons bounded together. Moreover, each pair of solitons is also in a bound-state consisting of two solitons. Then, they form double bound-state solitons as a whole unit. The temporal separation between pairs of bound-state solitons is 12 ps, well according with the 0.68 nm modulation spacing on the spectrum. It should be noted that the separation between the two solitons in each pair of bound-state is 1.81 ps, corresponding to the 4.52 nm modulation spacing on the spectrum. However, this modulation could not be clearly labeled on the spectrum, because it should coincide with the whole spectrum. Figure 2(d) gives the radio-frequency spectrum of the pulse. The signal-to-noise ratio (SNR) is ~52 dB, indicating the stable mode-locked operation of the fiber laser.

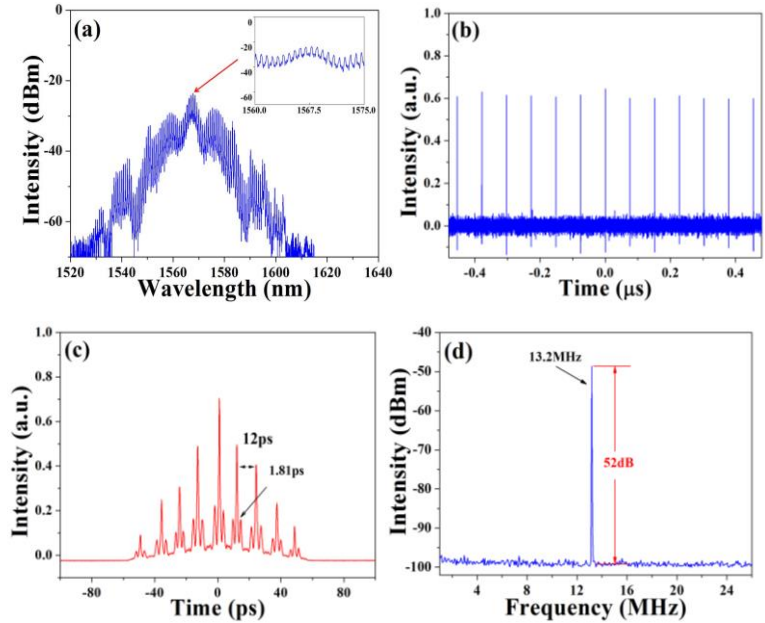

Fig. 2. The double bound-state solitons operation: (a) spectrum; (b) pulse-train; (c) autocorrelation trace; (d) radio-frequency spectrum.

Then the double bound-state solitons with 2.29 mW power were amplified by the EDFA and injected into the HNLF to generate supercontinuum. It is worth noting that the HNLF we adopted has a zero-dispersion wavelength (ZDW) at 1550 nm. So the wavelength longer than the ZDW operates at the anomalous dispersion regime and the wavelength shorter than the ZDW operates at the normal dispersion regime. By tuning the pump power of the EDFA, different stages of supercontinuum evolution were observed and summarized in Fig. 3. The whole excited spectrum finally covers the wavebands from 1100 nm to 2300 nm at the maximum pump power of 660 mW, with the 11-dB bandwidth reaching 1140 nm. Remarkably, there is no residual spectral peak at the pump wavelength.

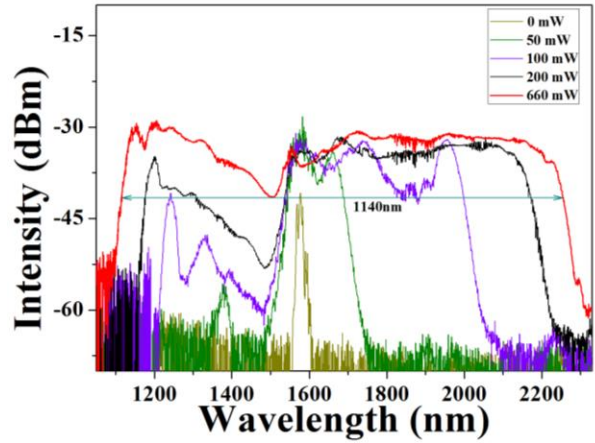

Fig. 3. Supercontinuum evolution pumped by double bound-state solitons versus the pump power.

Considering the pump wavelength and dispersion characteristic of the HNLF, the broadening process of the spectrum could be expounded by SPM, four-wave mixing (FWM), soliton fission and SRS. In longer wavebands, soliton fission causes soliton self-frequency shift (SSFS). Then the generated red-shift Raman solitons act as a secondary pump source for continuously exciting SRS, which makes the spectrum broaden greatly. Meanwhile, in shorter wavebands, owing to the red-shift Raman soliton and dispersion, the blue-shift dispersive wave is observed together with anti-Stokes lasing lines. Because of the influence of Raman amplification, the intensities of anti-Stokes components are lower than those of Stokes components stimulated in longer wavebands, which makes the asymmetric intensity distribution on both sides of pump wavelength in the early stage of spectral broadening. In subsequent stages, the generated dispersive wave keeps continuous blue shift, as presented in Fig. 3. This is because these dispersive waves can be further shifted by cross-phase modulation (XPM) initiated by infrared Raman soliton [24]. In the meantime, various FWM processes induced by SPM, XPM and ZDW where phase-matching is fulfilled generate new frequency components. Thus, the supercontinuum becomes flat and smooth. Noting that wavebands ranging from 1350 nm to 1420 nm and from 1800 nm to 1940 nm display strong oscillation structures, which is caused by peak absorption loss of water and carbon dioxide [25,26].

In particular, we notice that when we keep boosting pump power of the EDFA, the spectrum always experiences continuous broadening without any apparent residual high intensity spectral peak at pump wavelength as reported in references [10,13,15-17]. The intensities of some broadening components even exceed that of pump wavelength, indicating that strong nonlinear effects shift a large quantity of energy from pump pulses to new frequencies. We infer that it arises from the multiple coherent pump of double bound-state solitons, i.e., two sets of coherent pulse pump mechanism. The sub-set of pulses has fixed phase difference between two solitons. At the same time, this sub-set of pulses as a unit forms the main-set of bound-state pulses with five "solitons" through locking their phases. These two sets of coherent pump pulses simultaneously participate in the SCG. Therefore, the double bound-state solitons could induce stronger nonlinear effects and make the energy at the pump wavelength efficiently transfer to new generated frequencies. Consequently, the generated supercontinuum has no residual spectral peak at the pump wavelength. In addition, we increased the pulse number in the main-set bound-state from the seed pulse laser to generate supercontinuum. It is found that the characteristics of the supercontinuum keep almost the same as those in Fig. 3. It is strongly inferred that the flat supercontinuum without residual spectral peak at the pump wavelength is decided by multiple coherent pump rather than pump pulse numbers.

*3.2 Comparison with supercontinuums generated by conventional bound-state soliton and single soliton*

In order to highlight the elimination of the residual pump peak in supercontinuum generated by using the double bound-state solitons as pump seeds, we adopted another two different types of pulses as pump seeds for comparison. One is conventional bound-state soliton generated from the same seed pulse laser. The other is conventional single soliton produced by a mode-locked fiber laser with a similar structure as shown in Fig. 1. The spectra and autocorrelation traces of the pulses are presented in Fig. 4. Figures 4(a) and (b) show the typical characteristics of the conventional bound-state soliton, which consists of five sub-solitons with separation of 12 ps. Figures 4(c) and (d) are the representative features of conventional soliton with 1.43 ps pulse width. Both of the pulses are amplified by the same EDFA mentioned above. For better comparing the properties of the supercontinuum obtained with disparate seed pulses under the same conditions as far as possible, we set up the power of the pulses to be amplified to 100 mW. Then all of the pulses are separately transmitted into the same HNLF referred above. Finally, we get three different supercontinuums, which are shown in Fig. 5. As seen from Fig. 5, the supercontinuum pumped by the conventional soliton covers from 1200 nm to 2100 nm, including a deeper spectral defect in shorter wavebands. In longer wavebands, the broadened spectrum is narrower than the other two supercontinuums. More importantly, a strong residual spectral peak at the pump wavelength is presented with an exceeded intensity more than 20 dB, meaning a relatively low energy conversion in spectral broadening. Note that there is a CW component in the initial spectrum of the soliton, it may also contribute to the appearance of the spectral peak in supercontinuum. On the other hand, for the conventional bound-state soliton pumped SCG and double bound-state solitons pumped SCG, the two spectra have similar wavebands and intensities. The difference between them is that, for the conventional bound-state soliton pumped SCG, there also exists an obvious spectral peak at the pump wavelength with a relative intensity of around 15 dB. It implies that both of the pump pulses undergo similar nonlinear frequency shift, but there exists a distinct difference in SCG pumped by the double bound-state solitons. During the process of the SCG, for the conventional bound-state soliton, it is only one set of pump pulses, while for the double bound-state solitons, there are two sets of coherent pump pulses with fixed phase differences. In addition, we notice that on the spectrum of the conventional bound-state soliton, there is no CW component. But it still generates a distinct spectral peak in the excited supercontinuum. This suggests CW component in pulse spectrum is not the only ingredient responsible for the formation of the spectral peak in SCG.

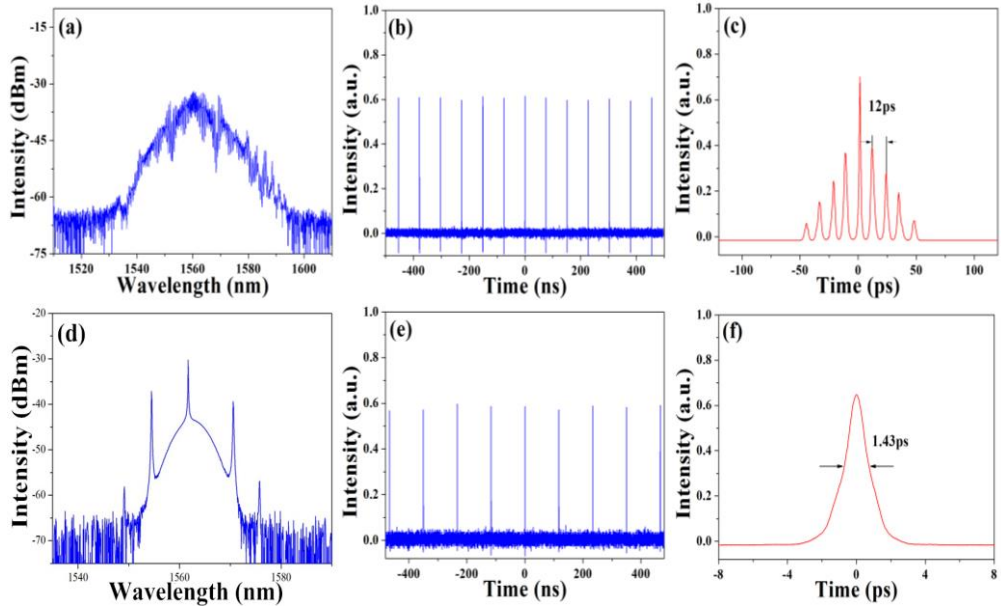

Fig. 4. Two types of seed pulses: (a) spectrum, (b) pulse-train and (c) autocorrelation trace of conventional bound-state soliton; (d) spectrum, (e) pulse-train and (f) autocorrelation trace of conventional soliton.

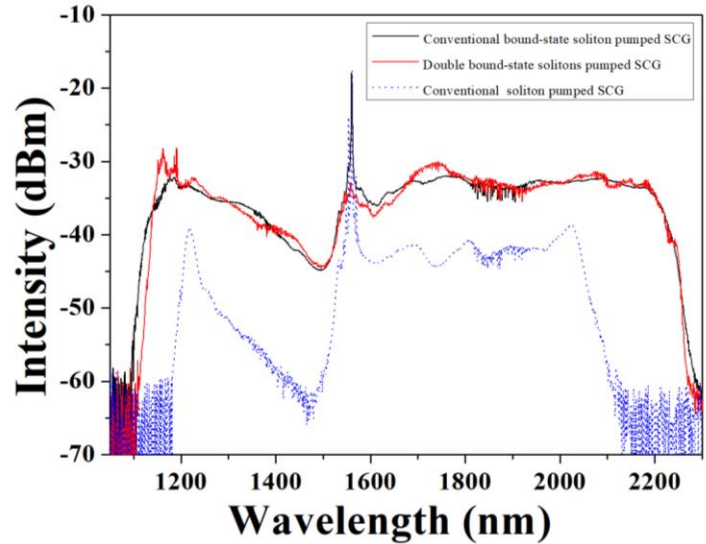

Fig. 5. SCGs obtained from different pump pulses.

In fact, according to the previous reports [20,27], the spectral peak at the pump wavelength in the supercontinuum generally occurs in two cases. One is CW or long pulse pumping (like Q-switched pulse or mode-locked nanosecond pulse). In this condition, modulation instability dominates the initial stage and generates low energy solitons with relatively low peak power, which leads to the input energy remain unconverted. Thus, the spectral peak emerges at the pump wavelength. The other is short pulse pumping (picosecond to femtosecond scale). In this condition, a part of the spectral peak comes from the original pump pulse, and the other is derived from the higher-order soliton compression in the initial stage, which will generate wide temporal pedestals at both wings of the pulse and most of the pulse energy is preserved in temporal pedestals [28]. Due to the low intensity of the temporal pedestals, it is unable to participate in nonlinear effects induced spectral broadening. In our experiment, we consider that the formation of the spectral peak in the generated supercontinuum pumped by conventional soliton and bound-state soliton arises from the inadequately converted seed pulse rather than soliton compression. Because in the case of the double bound-state solitons as the pump seeds, the pulses would also undergo higher-order soliton compression in the process of SCG due to the sub-pulses with narrower pulse widths. However, the generated supercontinuum is clear without a sharp spectral peak at the pump wavelength. Therefore we speculate that the higher-order soliton compression is not the reason for the spectral peak in the supercontinuum in Fig. 5. If the energy of the original seed pulse could be adequately converted to the new generated frequencies, the supercontinuum would be more flat without the sharp spectral peak. In our experiment, since the double-bound state solitons possess two sets of pulses with fixed phase differences, they could enhance the nonlinear effects in the process of generating supercontinuum by much stronger interactions among the pulses. Thus, the transferring efficiency of the pump energy could be greatly improved and effectively eliminate the residual spectral peak in the supercontinuum.

In addition, we notice that no matter what kinds of pump pulse we used, there always exists a spectral dent in the supercontinuum. Due to Raman amplification effect, as we know, the intensities of Stokes components are always much stronger than those of anti-Stokes light. Thus it makes the spectral collapse near the ZDW. Actually, when we continue to increase the pump power of the EDFA, the spectral dent is gradually improved, indicating that various FWM processes contribute to smooth the broadened spectra, as is shown in Fig. 3. Since the maximum pump power of the homemade EDFA is 660 mW, which brings a limit for the spectral dent. We believe if providing a higher pump power of the EDFA, the spectral dent will be further improved and the whole excited spectrum will be more flat and smooth.

### 4. Conclusion

In conclusion, we report for the first time, to the best of our knowledge, the residual pump peak as a universal phenomenon in SCG in previous reports is successfully eliminated through a novel multiple coherent pump scheme. The pump scheme is performed by double bound-state solitons from a mode-locked fiber laser. The supercontinuum covers from 1100 nm to 2300 nm with an 11-dB bandwidth of 1140 nm. Furthermore, we compare the SCG pumped by the conventional bound-state soliton and single soliton. The physical mechanism of the residual pump peak elimination owes to higher transferring efficiency of multiple coherent pump seeds. Our experimental results provide an interesting and novel technique for further researching on mechanism of SCG and improving the flatness of the supercontinuum.

### Funding

National Natural Science Foundation of China (Grant Nos. 61875058, 61875242, 11874018, 61378036); Science and Technology Program of Guangzhou (Grant No. 201607010245); Natural Science Foundation of Guangdong Province (No. 2018A030313347).

### References

1. T. Morioka, K. Mori, and M. Saruwatari, "More than 100-wavelength-channel picosecond optical pulse generation from single laser source using supercontinuum in optical fibres," Electron. Lett. **29**(10), 862-864 (1993).
2. A. Ruehl, M. J. Martin, K. C. Cossel, L. Chen, H. McKay, B. Thomas, C. Benko, L. Dong, J. M. Dudley, M. E. Fermann, I. Hartl, and J. Ye, "Ultrabroadband coherent supercontinuum frequency comb," Phys. Rev. A **84**, 011806 (2011).
3. C. Zhang, Y. Qiu, R. Zhu, Kenneth K. Y. Wong, and K. K. Tsia, "Serial time-encoded amplified microscopy (STEAM) based on a stabilized picosecond supercontinuum source," Opt. Express **19**(17), 15810-15816 (2011).

4. M. Maria, I. B. Gonzalo, T. Feuchter, M. Denninger, P. M. Moselund, L. Leick, O. Bang, and A. Podoleanu, "Q-switch-pumped supercontinuum for ultra-high resolution optical coherence tomography," Opt. Lett. **42**(22), 4744-4747 (2017).
5. R. R. Alfano, and S. L. Shapiro, "Emission in the region 4000 to 7000 Å Via four-photon coupling in glass," Phys. Rev. Lett. **24**(11), 584-587 (1970).
6. C. Lin, and R. H. Stolen, "New nanosecond continuum for excited-state spectroscopy," Appl. Phys. Lett. **28**(4), 216-218 (1976).
7. I. Ilev, H. Kumagai, K. Toyoda, and I. Koprinkov, "Highly efficient wideband continuum generation in a single mode optical fiber by powerful broadband laser pumping," Appl. Opt. **35**(15), 2548–2553 (1996).
8. N. Nishizawa and T. Goto, "Widely broadened super continuum generation using highly nonlinear dispersion shifted fibers and femtosecond fiber laser," Jpn. J. Appl. Phys. **40**, L365–L367 (2001).
9. A. V. Husakou and J. Herrmann, "Supercontinuum generation of higher-order solitons by fission in photonic crystal fibers," Phys. Rev. Lett. **87**(20), 203901 (2001).
10. D. D. Hudson, S. A. Dekker, E. C. Mägi, A. C. Judge, S. D. Jackson, E. Li, J. S. Sanghera, L. B. Shaw, I. D. Aggarwal, and B. J. Eggleton. "Octave spanning supercontinuum generation in an As2S3 taper using ultralow pump pulse energy," Opt. Lett. **36**(70), 1122-1124 (2011).
11. A. Zaytsev, C.-H. Lin, Y.-J. You, C.-C. Chung, C.-L. Wang, and C.-L. Pan, "Supercontinuum generation by noise-like pulses transmitted through normally dispersive standard single-mode fibers," Opt. Express **21**(13), 16056–16062 (2013).
12. C. Strutynski, P. Froidevaux, F. Désévédavy, J.-Ch. Jules, G. Gadret, A. Bendahmane, K. Tarnowski, B. Kibler, and F. Smektala, "Tailoring supercontinuum generation beyond 2 μm in step-index tellurite fibers," Opt. Lett. **42**(2), 247–250 (2017).
13. W. J. Wadsworth, N. Joly, J. C. Knight, T. A. Birks, F. Biancalana, and P. St.J. Russell, "Supercontinuum and four-wave mixing with Q-switched pulses in endlessly single-mode photonic crystal fibres," Opt. Express **12**(2), 299–309 (2004).
14. H. Chen, X. Zhou, S. P. Chen, Z. F. Jiang, J. Hou, "Ultra-compact Watt-level flat supercontinuum source pumped by noise-like pulse from an all fiber oscillator," Opt. Express **23**(26), 32909-32916 (2015).
15. N. Wang, J.-H. Cai, X. Qi, S.-P. Chen, L.-J. Yang, and J. Hou, "Ultraviolet-enhanced supercontinuum generation with a mode-locked Yb-doped fiber laser operating in dissipative-soliton-resonance region," Opt. Express **26**(2), 1689-1696 (2018).
16. J. Swiderski, M. Michalska, and P. Grzes, "Broadband and top-flat mid-infrared supercontinuum generation with 3.52 W time-averaged power in a ZBLAN fiber directly pumped by a 2-μm mode-locked fiber laser and amplifier," Applied Physics B **124**, 152 (2018).
17. S. Liang, L. Xu, Q. Fu, Y. Jung, D. P. Shepherd, D. J. Richardson, and S. Alam, "295-kW peak power picosecond pulses from a thulium-doped-fiber MOPA and the generation of watt-level >2.5-octave supercontinuum extending up to 5 μm," Opt. Express **26**(6), 6490-6498 (2018).
18. L. Yang, B. Zhang, T. Wu, Y. Zhao, and J. Hou, "Watt-level mid-infrared supercontinuum generation from 2.7 to 4.25 μm in an erbium-doped ZBLAN fiber with high slope efficiency," Opt. Lett. **43**(13), 3061–3064 (2018).
19. T. S. Saini, A. Kumar, R. K. Sinha, "Broadband mid-infrared supercontinuum spectra spanning 2–15 μm using $As_2Se_3$ chalcogenide glass triangular-core graded-index photonic crystal fiber," J. Lightwave Technol. **33**(18), 3914-3920 (2015).
20. T. Hori, J. Takayanagi, N. Nishizawa, and T. Goto, "Flatly broadened, wideband and low noise supercontinuum generation in highly nonlinear hybrid fiber," Opt. Express **12**(2), 317-324 (2004).
21. A. Kudlinski, G. Bouwmans, M. Douay, M. Taki, A. Mussot, "Dispersion-engineered photonic crystal fibers for CW-pumped supercontinuum sources," J. Lightwave Technol. **27**(11), 1556-1564 (2009).
22. D. Y. Tang, W. S. Man, H. Y. Tam, and P. D. Drummond, "Observation of bound states of solitons in a passively mode-locked fiber laser," Phys. Rev. A **64**, 033814 (2001).
23. P. Grelu, F. Belhache, F. Gutty, and J. M. Soto-Crespo, "Phase-locked soliton pairs in a stretched-pulse fiber laser," Opt. Lett. **27**(11), 966-968 (2002).
24. T. Schreiber, T. V. Andersen, D. Schimpf, J. Limpert, and A. Tünnermann, "Supercontinuum generation by femtosecond single and dual wavelength pumping in photonic crystal fibers with two zero dispersion wavelengths," Opt. Express **13**(23), 9556 (2005).
25. O. Humbach, H. Fabian, U. Grzesik, U. Haken, and W. Heitmann, "Analysis of OH absorption bands in synthetic silica," J. Non-Cryst. Solids **203**, 19-26 (1996).
26. A. Stark, L. Correia, M. Teichmann, S. Salewski, C. Larsen and V. M. Baev, "Intracavity absorption spectroscopy with thulium-doped fibre laser," Opt. Commun. **215**, 113-123 (2003).
27. S. M. Kobtsev, S. V. Kukarin, and S. V. Smirnov, "All-fiber high-energy supercontinuum pulse generator," Laser Phys. **20**(2), 375-378 (2010).
28. G. P. Agrawal, Applications of Nonlinear Fiber Optics, (Academic Press, San Diego, 2001).